# Coordinating Collaborative Chat in Massive Open Online Courses

Gaurav Singh Tomar, Sreecharan Sankaranarayanan, Xu Wang and Carolyn Penstein Rosé
Carnegie Mellon University, 5000 Forbes Avenue, Pittsburgh, Pennsylvania, USA – 15213
{gtomar, sreechas, xuwang, cprose}@cs.cmu.edu

**Abstract:** An earlier study of a collaborative chat intervention in a Massive Open Online Course (MOOC) identified negative effects on attrition stemming from a requirement for students to be matched with exactly one partner prior to beginning the activity. That study raised questions about how to orchestrate a collaborative chat intervention in a MOOC context in order to provide the benefit of synchronous social engagement without the coordination difficulties. In this paper we present a careful analysis of an intervention designed to overcome coordination difficulties by welcoming students into the chat on a rolling basis as they arrive rather than requiring them to be matched with a partner before beginning. The results suggest the most positive impact when experiencing a chat with exactly one partner rather than more or less. A qualitative analysis of the chat data reveals differential experiences between these configurations that suggests a potential explanation for the effect and raises questions for future research.

## Introduction

The field of Computer Supported Collaborative Learning (CSCL) has a rich history extending for nearly two decades, covering a broad spectrum of research related to learning in groups, especially in computer mediated environments. In contrast, a major limitation of the current generation of Massive Open Online Courses (MOOCs) is a lack of social presence. Analyses of attrition and learning in MOOCs both point to the importance of social engagement for motivational support and overcoming difficulties with material and course procedures. Effective collaborative learning experiences are known to provide many benefits to learners in terms of cognitive, metacognitive, and social impact (Kirschner, Paas, & Kirschner, 2009; Webb & Palinscar, 1996). These experiences offer a potentially valuable resource for MOOCs, if affordances can be provided that facilitate high quality collaborative learning interactions in the absence of human facilitators that can keep up with the high enrolment in such courses.

Traditionally, MOOC platforms, including team based MOOC platforms like NovoEd, have not offered synchronous interaction opportunities or even instantaneous forms of social awareness. MOOC contexts allow for asynchronous interaction and possibly even collaboration. However, the asynchronous nature leads to some unsavory experiences. For example, sometimes participants spend time posting a thoughtful post but get only a cursory reply. Participants sometimes have to wait days or even weeks to get a response to a question or receive feedback on completed work. This lack of immediacy implies that information can be out of date by the time someone views it, or worse, that the student needing help or feedback gave up and dropped out before the response was posted. These types of issues can hinder motivation. These limitations at the interface level stem from limitations at the architecture level due to challenges in scaling immediate update protocols. Through integration protocols such as the Learning Tools Interoperability protocol (LTI) chat tools and other forms of synchronous interaction have found their way into a few MOOCs. However, just the ability to integrate a synchronous chat room into a MOOC does not solve the problem. Challenges remain regarding coordinating the times of the discussions as well as supporting the functioning of an ongoing discussion.

The proposed solution to this problem is to support synchronous interaction within the MOOC context. Students may prefer synchronous activities because it offers them a greater experience of active involvement and social connection. Coordinating synchronous social engagement in a MOOC is challenging, however, and the challenges that arise can be frustrating for students. An earlier study identified positive and negative effects on attrition of a collaborative chat intervention that required students to be matched with exactly one partner prior to beginning the activity (Ferschke et al., 2015b). Negative effects occurred when students had to return multiple times in an attempt to be matched with a partner. That study raised questions about how to provide the benefit of synchronous social engagement without the coordination difficulties. In this paper we present a careful analysis of an intervention designed to overcome coordination difficulties by welcoming students into the chat on a rolling basis as they arrive rather than requiring them to be matched with a partner before beginning. This design raises questions about how the impact of the experience differs depending on conditions within the less controlled social environment within the chat.

In the remainder of the paper, we describe the motivation for the study from the Computer Supported Collaborative Learning (CSCL) literature. We then describe the intervention we evaluate in this paper. Next we present both quantitative and qualitative analyses of our intervention. Finally, we conclude with discussion and future directions.

## Theoretical Foundation

The fact that conversational interaction may provide an opportunity for (collaborative) learning has been underscored by many learning theories ranging from cognitive to socio-cultural perspectives (Kirschner, Paas, & Kirschner, 2009; Webb & Palinscar, 1996). However, it is well known that without support, many instances of collaborative learning fail. In light of this fact, the field of collaborative learning has produced a wide variety of forms of scaffolding often referred to as scripts. Collaboration scripts may operate at the macro level, providing task structuring and role assignment. Or it may operate at the micro level, structuring the nature of the flow of contributions to the discourse.

Our goal is to find the best ways to provide learners with a space to interactively hone their understanding of concepts related to a specific domain so that they have the chance to display their own reasoning, experience how others display their reasoning, challenge and be challenged by others. In order for the interactions to provide a meaningful learning experience for students, it was essential for these activities to be well integrated into the instructional design, and not treated as an afterthought or an appendage. This leads to more authentic, and ecologically valid learning experiences for students (Tudor, 2000; Van Lier, 2004). In addition to this embedding, interactions are mainly meaningful for learning when they are structured and scaffolded in an appropriate manner.

In this paper, we build on a paradigm for dynamic support for group learning that has proven effective for improving interaction and learning in a series of online group learning studies conducted in classroom settings. In particular we refer to using tutorial dialogue agent technology to provide interactive support within a synchronous collaborative chat environment (Kumar et al., 2007; Chaudhuri et al., 2008; Chaudhuri et al., 2009; Kumar et al., 2010; Ai et al., 2010; Kumar & Rosé, 2011). Introduction of such technology in a classroom setting has consistently led to significant improvements in student learning (Adamson et al., 2014), and even positive impacts on the classroom environment outside of the collaborative activities (Clarke et al., 2013). While it would seem to be desirable to import such technology into a MOOC setting to provide a learning experience that is both more instructionally valuable and socially supportive, such an introduction comes with its technical, methodological, and theoretical challenges.

In our setup, the goal is for the agent to aid the students in integrating their respective understandings of the concepts they previously encountered individually in the video lectures and other assignments that precede the chat activity. On the other hand, it also provides an opportunity to develop their collaborative skills (O'Donnell, 1999).

## Collaboration Platform

The chat tool system deployed in the MOOC study reported in this paper is Bazaar (Adamson et al., 2014), which is used to provide interactive support within online synchronous collaborative chat activities. Reflection activities were authored by the instructor team to provide students the opportunity to reflect substantively about the material they learned in each unit. Each chat activity included several specific reflection topics to provide a consistent macro level structure to the chats for each unit. Very little micro-level support was offered in addition to the macro-level structuring.

The most similar previously published study regarding a collaborative chat intervention in a MOOC was published by Ferschke and colleagues (Ferschke et al., 2015a). In that study, in order to facilitate the formation of ad-hoc study groups for the chat activity, Ferschke and colleagues (Ferschke et al., 2015b) made use of a simple setup referred to as a Lobby. Students entered the Lobby with a simple, clearly labeled button integrated with the edX platform. Upon entering the Lobby, students were asked to enter a username that would be displayed in the chat. Once registered in the Lobby, the student waited to be matched with another participant. If the student was successfully matched with another learner who arrived at the Lobby within a couple of minutes to interact with, he and his partner were then presented with a link to click on to enter a chat room created for them in real time. Otherwise they were requested to come back later. Some students needed to make up to 15 attempts in order to be successfully matched for a chat. Thus, many students were frustrated. A follow up analysis (Ferschke et al., 2015b) reports a negative impact on commitment to the course for students who experienced this frustration. An important lesson learned from this study was that whereas providing the opportunity for synchronous chat was positive for students for whom it was possible to be matched for a chat easily, this positive effect was balanced with a negative effect in the case where the lack of critical mass despite the total enrollment of 20,000 students from their MOOC was not sufficient to enable a quick match.

In order to address the difficulties reported in the earlier study (Ferschke et al., 2015b), we did not employ an intermediate Lobby interface in our design. Instead we exposed a single, continuous chat room that multiple students could join at any time on a rolling basis, rather than beginning all at the same time, and without having to be matched with a particular peer student. The conversation continues as long as there is at least one person in the room. The architecture is able to facilitate a chat with only one student present and keeps the conversation going for any number of participants. That way, students can enter the chat on their own schedule and join other students who are engaged in the activity. The chat room automatically resets once all participants have left the chat.

In order to provide a very light form of micro-level script-based support for the collaboration, the computer agent facilitator in our design kept track of the students in the room and prompts that had been given to the students. For each student, it tracked the events of their joining and leaving the chat room. For topic prompts, it tracked the times they were given and the students who were present in the chat room at that time. This tracking was done to make sure each student could engage with all question prompts of the reflection exercise and no student saw the same question prompt twice unless the activity script restarted and students decided to repeat it.

Another design consideration was to address a difficulty identified in the earlier study, which was that the students did not always stay on topic and sometimes engaged in goalless interactions or neglected to interact at all. One possible explanation is that these students may have misunderstood the question. To address this issue in the current design, the agent kept track of any student and group inactivity that went on for more than two minutes. It also kept track of how many utterances each student contributed. It aggregated students' messages and computed a similarity score between student messages and the topic prompt to test whether the students were on topic. This was also done over a two-minute window. If the group was dormant or did not say anything that is relevant to the topic in that two-minute window then the agent generated a poke message, which rephrased the question again by providing timely hints that make it more explicit what is expected from a contribution. A contentful group discussion on a question prompt was observed to last for about 10 minutes on average.

Although a single, continuous chat room provides the capacity to solve the problems with synchronous collaboration in a MOOC observed in the earlier study, it may also entail additional coordination challenges. For instance, a student may not be motivated to join in the middle of the discussion or might feel lost due to a lack of a frame of reference for where to start upon joining an in progress discussion. To address this issue, we introduced a prompt for the agent to ask the group to summarize the state of the conversation for incoming students in order to help ramp up new participants and encourage current participants to voice their understanding of discussed concepts. Alternatively, sometimes an agent-generated summary is given, mentions a list of topics already discussed as well as a summary of the current topic. To ensure the agent does not interrupt the discussion with repeated summarization requests, these requests were issued only if at least two topic prompts had been discussed.

## Results

The most positive result of the earlier Ferschke et al. study (Ferschke et al. 2015b) was the observed reduction in attrition associated with participation in a collaborative chat as measured through a survival analysis. For our work, it is important to determine not only whether we have accomplished the specific changes to student experiences that motivated our design work. We also want to ensure that we maintain or even enhance the positive impact observed with the earlier approach. In order to measure the impact of participation in a collaborative chat on attrition, we adopt a survival analysis approach. Survival analysis is a statistical modeling technique used to model the effect of one or more indicator variables at a time point on the probability of an event occurring on the next time point. In our case, we are modeling the effect of participation in a collaborative chat on probability that a student ceases to participate actively in the course on the next time point.

### Methodology

Survival models are a form of proportional odds logistic regression, and they are known to provide less biased estimates than simpler techniques (e.g., standard least squares linear regression) that do not take into account the potentially truncated nature of time-to-event data (e.g., users who had not yet ceased their participation at the time of the analysis but might at some point subsequently). In a survival model, a prediction about the probability of an event occurring is made at each time point based on the presence of some set of predictors. The estimated weights on the predictors are referred to as hazard ratios. The hazard ratio of a predictor indicates how the relative likelihood of the failure (in our case, student dropout) occurring increases or decreases with an increase or decrease in the associated predictor in the case of a continuous variable, or presence vs. absence of the factor in the case of a binary variable. A hazard ratio of 1 means the factor has no effect.

If the hazard ratio is a fraction, then the factor decreases the probability of the event. For example, if the hazard ratio was a number of value .4, it would mean that for every standard deviation greater than the average of the continuous predictor variable (or where a data point has value 1 for a binary predictor variable), the event is 60% less likely to occur (i.e., 1 - n). If the hazard ratio is instead greater than 1, that would mean that the factor has a positive effect on the probability of the event. In particular, if the hazard ratio is 1.25, then for every standard deviation greater than the average of the continuous predictor variable (or where the data point has value 1 for a binary predictor variable), the event is 25% more likely to occur (i.e., n - 1).

Survival analyses are correlational analyses, and as such they do not provide causal evidence for an effect. However, lack of an effect in a survival analysis would suggest that the data fail to provide causal evidence as well. A positive effect in a survival analysis would suggest that it makes sense as a next step to manipulate the associated factor so that causal evidence for a positive effect could be measured.

## Specifying the Model

In our survival model we include control variables, independent variables, and a dependent variable. Our primary interest is how the independent variables related to participation in collaborative chats make predictions about the dependent variable, which indicates course dropout. However, control variables are essential in accounting for variances in the participants that may influence attrition. For example, some students are more active in the course in general, indicating a priori greater commitment to the course. If we do not account for this in our model, we cannot say whether it's the intervention we are introducing that is leading to the observed difference in the dependent variable, or it's because other confounding variables such as the difference priori commitment level of students' drives their participation in the intervention that leads to the effect.

**Unit of analysis.** In order to assess the impact of measured factors at each time point during a student's trajectory through the course, it is necessary to decide what the unit of analysis is. In other words, it is necessary to determine what the time interval to use in the survival model is. Even the most active participants in the course did not participate every day. However, very active participants returned to the course more than once within a week. Based on preliminary exploration into the dataset, we chose our unit of analysis to be a 7-day period of time.

**Student population.** We conducted our analysis in an edX MOOC called Big Data in Education (BDEMOOC), which was taught by Ryan Baker from Teacher's College, Columbia University. The MOOC was launched in the fall of 2015.

We included in our analysis all students who had at least clicked once during the course to enter a chat. There were 401 such students in our dataset. The collaborative chat activities were positioned as enrichment activities after the individual work for the week was completed. In order to enter a chat, students clicked on a button in the courseware page. We expect that students who attempted to participate in a chat were more active on average than students who did not since the chat activities were positioned at the end of the unit, so they would be mainly consumed by students who had completed the other assignments of the course. However, students who did not complete the other activities were not prevented from participating. The baseline data points in our model are the ones when students did not click to enter a chat during the corresponding time interval.

**Control variables.** An important indicator of a priori commitment to active engagement was the number of clicks on course videos. We included a standardized count of video clicks during the time period between two consecutive data points as a control variable in our analysis.

**Independent variables.** As mentioned, the baseline for our comparison is data points that did not include any attempt to enter into a chat. There were 21,968 of these. If a student clicked to enter a chat, four possible things could happen. 1) Some number of students experienced technical difficulties due to their Internet configuration, which didn't allow them to connect to the chat server port from their network. There were 145 such data points, which we label as Malfunction data points. 2) Another possibility is that the student entered the chat, but no other student joined them there. There were 349 such data points. We label these as alone data points. 3) There were 145 data points where students entered a chat and there was exactly one partner student. We labeled these as Pair data points. 4) Finally there were 15 data points where there were more than two students present. We refer to these as Group data points. We included binary independent variables in our survival model for Malfunction, Alone, Pair, and Group.

In order to assign these binary variables, it was necessary to compute the maximum number of peers that the student interacted with in a particular chat session. We calculated this number by counting the number of peers present each time the student performs an event (join, leave, post a message). From the set computed during a chat session we take the maximum. This number reflects the largest group size the student can possibly have actively interacted with during the session.

**Dependent variable.** We referred to the dependent variable as Drop. This was a binary variable that was 1 for the last time interval of a student's active participation in the MOOC and 0 otherwise.

## Survival Model Results

The main results of the analysis (summarized in Table 1) are consistent with those of the earlier study (Ferschke et al., 2015b) in that we see a consistent trend suggestive of positive impact when students participated in a chat. The most positive impact is in the case where students chat with exactly one partner student. There the hazard ratio is .6 indicating a 40% increase in probability that the student will still be active at the next time point. However, the effect is only marginal (p= .06). The trend for data points where students chat alone or with more than one partner student is also positive, however the effect is not significant. There is a significant negative impact of experiencing a malfunction due to network connectivity issues. However, even this is less strong than that reported by (Ferschke et al., 2015b) for students who were required to click multiple times in order to be matched for a chat. Ferschke et al. (2015b) report a hazard ratio of 2.33 for a standardized count of match attempts. Overall, the results suggest that welcoming students into the chat on a rolling basis rather than requiring them to be matched with a partner student in order to enter is a more advantageous strategy since students who experience a pair chat benefit, and those who chat alone or with more than one student are not harmed.

Table 1. Survival table with estimates that measure the impact of control variable (Video Clicks), and independent variables (Malfunction, Alone, Pair, and Group) on probability of survival

| Independent Variable | Hazard Ratio | p-value |
| --- | --- | --- |
| Standardized Video Clicks | .97 | p= n.s. |
| Malfunction | 1.7 | p < .001 |
| Alone | .89 | p= n.s. |
| Pair | .6 | p =.06 |
| Group | .8 | p = n.s. |

## Taking a Closer Look: Post Hoc Analyses of Results

The results of the survival model suggest that students who participated in the BDEMOOC and chatted with exactly one partner student experienced more benefit. In our Post Hoc analysis we attempted to better understand what was different in their experience. We find evidence that students experienced a richer interaction when they participated in the chat with a single partner student. One strong indicator of this richness is in comparing amount of time spent in the chat based on how many students were participating. We see a significant difference in the average time spent in the chat between students who had a partner in the chat and student who didn't ($F(2,563) = 3.1, p < .05$). Chats with more than two students were not significantly longer or shorter in the average time of the session than either. Informally, we also see less frustration and more exchanged reasoning in the chats containing exactly two students.

Table 2. Comparison of average time of chat sessions in relation to the number of students participating.

| Number of students in the chat room | Mean time spent | Standard deviation | Median time spent |
|---|---|---|---|
| 1 | 389.44 | 755.67 | 0.0 |
| 2 | 501.58 | 874.23 | 1920.0 |
| 3 | 582.86 | 586.31 | 1560.0 |

In order to further investigate what is the difference between dialogues in chat rooms that contain only one person, two people, and more than two people, we manually coded frustration in the dialogues and compared the difference between different conditions. Since frustration was a factor that influenced dropout in the Ferschke et al. (2015b) study, we believed this would offer a useful lens on the chat processes at work in our study.

Two coders coded 100 dialogue turns generated by learners in the collaborative chats as expressing frustration or not in order to check the inter rater reliability of frustration coding. The coders achieved a kappa of 0.784. The two coders then continued to code the rest of the dataset, which contains 2917 dialogue turns in total. The coders then computed the average number of contributions coded as expressing frustration in each condition. The trend in proportion of contributions expressing frustration across sets was consistent with expectations. The proportion of contributions expressing frustration was lowest in the chats with exactly two participants. The rooms with single student generate the highest concentration of messages expressing frustration (5.78%), and the rooms with more than two students rank second (3.59%), while the rooms with pairs generate the lowest concentration of messages expressing frustration (2.86%). In testing the significance of the difference between the three conditions, we found that only the difference in the proportion of frustration messages expressed between the condition of single students and that of pairs is marginally significant (Z-score = 1.7662, p= 0.077).

We then took a closer look at what happened in the chat room in different conditions. When there is only one person in the chat room, the student tends to check whether others are present, and sometimes expresses boredom and disappointment at being alone. Here are examples from several different chats with individual students:

*Student 1:* shrugs shoulders
*Student 2:* Is it just me and VirtualRyan?
*Student 3:* Am I missing anything? I feel silly just chatting into space! :P

Another student was excited in the beginning, and he talked to himself for a while, nobody responded to him, and then he left.

*Student 4:* Is anyone here?
Anyone wants to discuss learning curves and other visualization methods?
This was the best week so far. The discussion about the various curves gave insight into how these techniques can be use to understand the impact of curriculum or how to help students directly.
I'll try again another time.
leave

Above we reported that students in pairs remained in the chat room longer than individual students. Informally we observed a higher concentration of expressed reasoning in the chats with pairs. Here are some examples of conversation when exactly two people are in the room. The examples below show that when there are two people in the chat room, they frequently exchanged perspectives, and offered positive comments on each other's ideas. We underlined the parts that they are exchanging, referring to, building upon and commenting on each other's ideas.

*Example 1:*
*P:* The dataset I looked at included more than time-related variables, so I would think yes. There are grades received at different stages, reason why a student took a class, etc.
*D:* Hmm, I agree with the baseline idea that the decision tree works better at categorical data.
*D:* Good. Thanks for the inspiration.

*Example 2:*
*T:* Well, to continue our theme, I would use data mining to predict MOOC dropout in China
*E:* That sounds interesting. What type of predictors were you thinking you may use?
*T:* I am going to use the "assignment accomplishment" predictor.
*T:* The reason I chose the predictor is based on my own experience. So how about yours?
*E:* I am working to develop new online programs and have access to all of the data in our LMS so thought I may be able to help tell students what they should be doing to increase their chance of success.
*E:* It may also help us check use of the LMS and flag students who may have persistence issues....just a thought. I don't have experience in data mining before.
*T:* And I would like to add a second feature: some complaints in the discussion area, to predict the dropout, as well.
*E:* Good thoughts! What all is included in the assignment accomplishment predictor?
*T:* Thanks. but just a thought, not very detail. I believe the more assignment accomplished, the more successful the subject may reach the endline.

## Discussion

In this paper, we presented an analysis of results from a collaborative chat intervention that suggests positive impact of the intervention when students experienced the chat with exactly one partner. A post hoc analysis suggests that students had a richer interaction when they were in the chat with exactly one partner. One possible explanation is that the support offered to groups was not sufficient to scaffold the interaction for more than two students. However, the number of chats with more than two students was small. And our analysis is merely correlational. Thus, we are unable to make strong claims about the reason for the pattern we found with this study.

One limitation of the current study is that some students experienced a negative impact from technical difficulties, namely the port connection issues discussed earlier. Though we carefully isolated the data points directly impacted by this issue in our analysis, we cannot eliminate the possibility that the negative impact of this issue on the students who directly experienced it may have affected others through hearing about the issues in the discussion forums. This issue has now been resolved in our infrastructure.

More substantively, despite the positive impact of collaborative chats when the conditions were ideal (e.g., where exactly one partner student was present), we identified some challenges with the intervention. For example, we observed that many students who were chatting with agent in the absence of peers felt that agent was not a good listener, felt bored, and sometimes left room in the middle of the discussion. They expressed a desire to be able to discuss the questions with the agent instead of just giving their reflections on the question prompt being asked. Another issue was that some dynamics of the agent were tuned to the situation where multiple human students were present and thus were not appropriate in the case where only one student was present. Thus, a more dynamic timing strategy for the task structuring may have been received better.

In our follow up work, in order to alleviate the boredom felt by individual students in the chat room, we are developing a design that supports a more substantive exchange between a student and the agent in the absence of other human partners in the interaction. Specifically, we will make use of tutorial dialogue agent technology, such as Knowledge Construction Dialogues (KCDs) (Rosé & VanLehn, 2005; Kumar & Rosé, 2011), which may provide more intensive engagement. Thus, if a single student is present in the room, the facilitation agent may lead the student step by step to construct substantive explanations related to the issues targeted by the collaborative chat exercise. Adaptive time of task structuring will also be employed.

## Conclusions

This paper addresses questions on how to effectively orchestrate synchronous discussion activities in MOOCs despite practical coordination challenges. It presents an evaluation of a specific collaborative chat intervention integrated within an edX MOOC. While an earlier study offered results associating participation in a chat session with reductions in attrition over time (Ferschke et al., 2015b), that study was not able to separate the effect of the chat activity from the effect of synchronous interaction with a peer. In this paper, we present a correlational analysis suggesting differential effects of a chat experience depending upon the number of peers present. The positive impact associated with interaction with a single partner student is consistent with the results presented in the earlier study, though slightly weaker, and thus provide more confidence in the finding that interaction with a peer is associated with at least temporary elevation in commitment to the course. The study also suggests further directions to improve the intervention for future MOOCs.

## Acknowledgement

This work was funded in part by grants from Google and the Gates foundation and NSF grant ACI-1443068.